\documentclass[sigconf]{acmart}

\copyrightyear{2023}
\acmYear{2023}
\setcopyright{rightsretained}
\acmConference[SIGIR '23]{Proceedings of the 46th International ACM SIGIR Conference on Research and Development in Information Retrieval}{July 23--27, 2023}{Taipei, Taiwan}
\acmBooktitle{Proceedings of the 46th International ACM SIGIR Conference on Research and Development in Information Retrieval (SIGIR '23), July 23--27, 2023, Taipei, Taiwan}\acmDOI{10.1145/3539618.3591943}
\acmISBN{978-1-4503-9408-6/23/07}

\makeatletter
\gdef\@copyrightpermission{
  \begin{minipage}{0.3\columnwidth}
   \href{https://creativecommons.org/licenses/by/4.0/}{\includegraphics[width=0.90\textwidth]{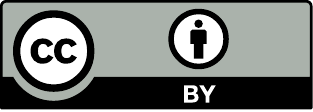}}
  \end{minipage}\hfill
  \begin{minipage}{0.7\columnwidth}
   \href{https://creativecommons.org/licenses/by/4.0/}{This work is licensed under a Creative Commons Attribution International 4.0 License.}
  \end{minipage}
  \vspace{5pt}
}
\makeatother
\usepackage{enumitem}
\usepackage{booktabs}
\usepackage{array,multirow,graphicx}
\usepackage{balance}

\begin{document}

\title{Adapting Learned Sparse Retrieval for Long Documents}

\author{Thong Nguyen}
\email{t.nguyen2@uva.nl}
\orcid{1234-5678-9012}
\affiliation{%
  \institution{University of Amsterdam}
  \city{Amsterdam}
  \country{Netherlands}
}

\author{Sean MacAvaney}
\email{sean.macavaney@glasgow.ac.uk}
\orcid{0000-0002-8914-2659}
\affiliation{%
  \institution{University of Glassgow}
  \city{Glasgow}
  \country{Scotland, UK}}

\author{Andrew Yates}
\email{a.c.yates@uva.nl}
\orcid{0000-0002-5970-880X}
\affiliation{%
  \institution{University of Amsterdam}
  \city{Amsterdam}
  \country{Netherlands}
}
\renewcommand{\shortauthors}{Thong Nguyen, Sean MacAvaney, \& Andrew Yates}

\begin{abstract}
Learned sparse retrieval (LSR) is a family of neural retrieval methods that transform queries and documents into sparse weight vectors aligned with a vocabulary. While LSR approaches like Splade work well for short passages, it is unclear how well they handle longer documents. We investigate existing aggregation approaches for adapting LSR to longer documents and find that proximal scoring is crucial for LSR to handle long documents. To leverage this property, we proposed two adaptations of the Sequential Dependence Model (SDM) to LSR: ExactSDM and SoftSDM. ExactSDM assumes only exact query term dependence, while SoftSDM uses potential functions that model the dependence of query terms and their expansion terms (i.e., terms identified using a transformer's masked language modeling head).

Experiments on the MSMARCO Document and TREC Robust04 datasets demonstrate that both ExactSDM and SoftSDM outperform existing LSR aggregation approaches for different document length constraints. Surprisingly, SoftSDM does not provide any performance benefits over ExactSDM. This suggests that soft proximity matching is not necessary for modeling term dependence in LSR.
Overall, this study provides insights into handling long documents with LSR, proposing adaptations that improve its performance.

\vspace{0.4em}
\hspace{5.5em}\includegraphics[width=1.25em,height=1.25em]{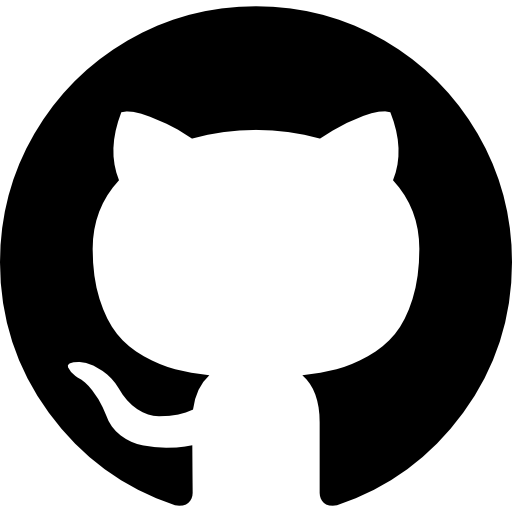}\hspace{.3em}
\parbox[c]{\columnwidth}
{
    \vspace{-.55em}
    \href{https://github.com/thongnt99/lsr-long}{\nolinkurl{github.com/thongnt99/lsr-long}}
}
\vspace{-1.5em}

% This work investigates the effectiveness of existing aggregation approaches for LSR to handle longer documents. The experiments reveal that most aggregation strategies exhibit a decreasing trend in effectiveness as document length increases, except for the max score aggregation. This observation suggests that proximal/local scoring is crucial for LSR to handle long documents. To leverage this property, two adaptations of the Sequential Dependence Model (SDM) to LSR are proposed: ExactSDM and SoftSDM. The former assumes only exact query term dependence, while the latter introduces potential functions that model the dependence of query terms and their synonyms.
% Our experiments on MSMARCO Document and TREC Robust04 datasets demonstrate that both ExactSDM and SoftSDM consistently outperform existing LSR aggregation approaches across different length constraints. However, SoftSDM shows no benefit over the ExactSDM, hinting that soft phrase/proximity matching is not necessary for modeling term dependence in LSR. 
\end{abstract}





\maketitle

\section{Introduction}
\begin{figure}
    \centering
    \includegraphics[width=\linewidth]{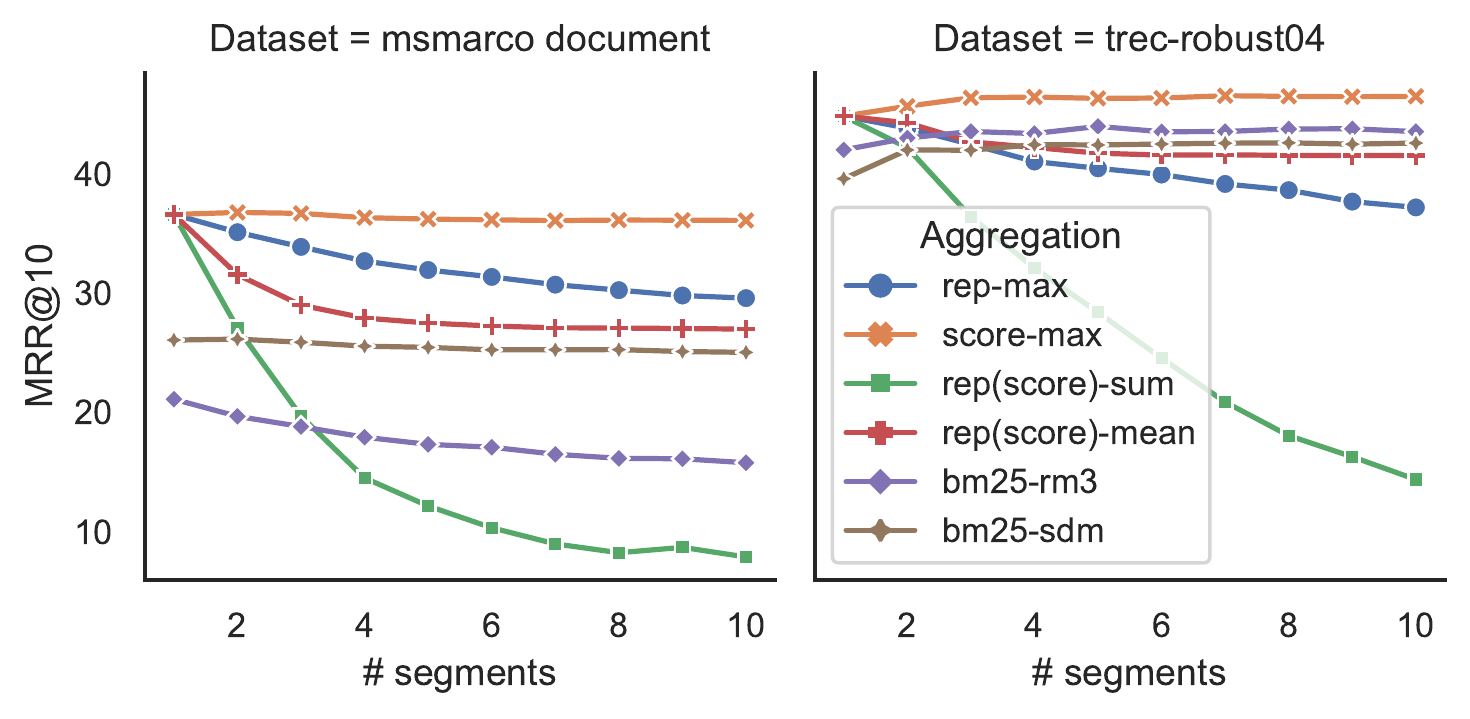}
    \caption{Performance of aggregation methods proposed for neural rankers with LSR and long documents.}
    \label{fig:lsr_long_figure}
    \vspace{-0.6cm}
\end{figure}

Learned sparse retrieval (LSR) is a recent family of first-stage information retrieval that utilizes neural models, typically transformers, to encode queries and documents into sparse weight vectors aligned with a vocabulary. By operating in the lexical space, LSR provides several benefits over dense retrieval, including greater transparency
%and generalizability to out-of-domain distributions
% AY: I wasn't sure what to cite for ^ 
while remaining comparably effective~\cite{formal2022distillation, formal2021splade, nguyen2023unified}. 
% \ay{What can we cite to support this?}
Moreover, LSR's compatibility with an inverted index allows for the reuse of techniques previously developed for lexical retrieval (e.g., BM25) ~\cite{lin2022proposed}. 
% However, some techniques, like query processing algorithms \cite{mallia2022faster, mackenzie2022accelerating, mackenzie2021wacky}, require modification to handle the change in term weight distribution.
% However, this approach poses a challenge in handling long documents due to the noise and content diversity. Multiple topics/aspects may present in a document, and only one of them may be relevant to the query.

Previous evaluations of LSR methods have focused on short texts due to transformers' input length limit, such as by focusing on the MSMARCO passage collection~\cite{msmarco} or truncating texts to the transformer's maximum input length.
%Despite the effectiveness of LSR, the input length limit of transformers has resulted in the evaluation of LSR methods on short texts \cite{msmarco} or texts truncated to the model's maximum length. 

To handle long documents, a common practice is to split the input into multiple segments, encode these segments individually, and aggregate the output. 
Several works have investigated score and representation aggregation strategies for the Cross-Encoder architecture, finding that representation aggregation typically outperforms score aggregation (e.g., taking the max of segments' vector representations rather than taking the max segment score)~\cite{macavaney:sigir2019-cedr,li2020parade,zhang2021comparing,boytsov2022understanding}.

We hypothesize that these insights may change for LSR due to the inherent difference in the encoding mechanism between Cross-Encoders and LSR. While Cross-Encoders encode the query and document simultaneously, LSR encodes queries and documents separately, which makes it challenging to judge what from a document segment is  not relevant to the information need.
As a result, the aggregation operation may accumulate noise and make documents less separable. 
In addition, as the documents get longer, it is particularly difficult for LSR, which relies solely on term-based representations, to deal with scattered term matches. Intuitively, matches within neighbouring text should be a stronger relevance signal than isolated matches present in different segments~\cite{Tao2007AnEO}.

In this work, we first study the effectiveness of existing aggregation approaches for adapting LSR to long documents. Specifically, we investigate three aggregation operators (\textit{max, sum, mean}) on two output levels (\textit{representation, score}). By applying these approaches to the state-of-the-art LSR method~\cite{formal2021splade,nguyen2023unified} on two datasets (MSMARCO Document, TREC Robust04), we find that most of the approaches, except for the max score aggregation, are fragile with long documents; as can be seen in Figure \ref{fig:lsr_long_figure}, their performance drops severely as more document segments are added into the representations. Max score aggregation, however, avoids this downward trend with mostly stable performance over different numbers of segments. This suggests that the proximity matching that inherently happens with max score aggregation is crucial for LSR to handle long documents.

In order to better utilize local proximity, we propose two approaches, ExactSDM and SoftSDM, that adapt the SDM~\cite{metzler2005modeling} for use with LSR. ExactSDM assumes a phrase or proximity match if the exact phrase or constituent terms appear in the document within a local document window. SoftSDM relaxes this assumption by allowing constituent terms to be softly matched with their expansion terms, as the Splade method does for unigrams.

We evaluate the performance of both ExactSDM and SoftSDM on the MSMARCO Document and TREC Robust04 datasets, finding that both approaches consistently outperform previous aggregation methods in the context of LSR. However, SoftSDM does not provide significant advantages over ExactSDM, making ExactSDM a more attractive option as it can be applied to a wider range of LSR methods, including both MLM-based architectures that leverage a masked language modeling (MLM) head to identify and score expansion terms ~\cite{formal2021splade,formal2022distillation, macavaney2020expansion} and MLP-based architectures that estimate term salience using a multilayer perceptron (MLP) with no soft matching via expansion terms~\cite{lin2021few, dai2019context, mallia2021learning}.

Overall, our work sheds light on adapting LSR to long documents, with an approach adapted from SDM that significantly outperforms existing aggregation techniques for neural retrieval methods.

\section{Background}
\subsection{Learned sparse retrieval}
Learned Sparse Retrieval (LSR) methods score a query-document pair by taking the dot product between their sparse vectors produced by a query encoder ($f_Q$) and a document encoder ($f_D$), i.e., $score(q,d) = f_{Q}(q) \cdot f_{D}(d) = w_q \cdot w_d = \sum_{i=1}^{|V|} w_q^i w_d^i$, where $w_i$ represents the predicted weight of the $i^{th}$ term in the vocabulary.

Compared to dense retrieval, LSR's vectors are high dimensional and sparse, with most of their elements being zero. The dimensions of these vectors are tied to a vocabulary, so LSR is closely connected to traditional sparse retrieval methods such as BM25. However, unlike BM25, LSR learns the term weights instead of obtaining them from corpus statistics like TF-IDF. This similarity with BM25 makes LSR compatible with many existing techniques such as the inverted index, which were previously built for purely lexical retrieval models like BM25~\cite{lin2022proposed}.

Splade is a state-of-the-art LSR method that uses a masked language modeling (MLM) head based on BERT to perform term expansion and term weighting end-to-end. Given an input query/document, Splade encodes it into a logit matrix $\mathbf{W}$, where $W_{i,j}$ represents the translation probability score
% \ay{translation score, maybe? it wouldn't be interpretable as a probability before softmax}
from the $i^{th}$ term in the input sequence to the $j^{th}$ vocabulary item ($|V| \approx 30k$). Splade then outputs, for each term in the vocabulary, a non-negative log-scaled weight, which is the maximum logit value of the term across the sequence, i.e., $w_k = \max_i(\log(1+\text{ReLU}(W_{i,k})))$. This max aggregation retains only the term weights and, like all LSR methods, it drops positional information. In other words, Splade uses term positional information when estimating query and document term weights, but is position-agnostic when scoring documents. Positional information is critical for inferring phrase semantics, however; \textit{``MU defeated Arsenal''} has the opposite meaning of \textit{``Arsenal defeated MU''}. Furthermore, IR Axioms suggest that documents that contain phrases from the query (or query terms in close proximity) are more relevant than those with query terms scatted throughout~\cite{Tao2007AnEO}.

% Term proximity concerns the distance between two or more matches, where matches within a local region should be more relevant than matches scattered in different places.

% seem irrelevant, could be removed later if we need space
% However, the learned weights also cause a distribution shift from the tf-idf scores, making some original query processing algorithms not very efficient for LSR. Several modifications have been proposed to resolve this issue \cite{citation, ciation}.
% 
% possibly move to introduction/motivation

% Term proximity is important for matching long documents. However, LSR are mostly evaluated on the MSMARCO passages which are generally short.

\subsection{Term Dependence Model} 
In prior work, researchers introduced techniques that incorporate the dependence of two or more terms into traditional bag-of-words retrieval models~\cite{huston2014comparison, metzler2005modeling, bendersky2010learning}. Among these techniques, the Sequential Dependence Model~\cite{metzler2005modeling} (SDM) is one of the most widely known and has been shown to be effective. 

SDM assumes a dependence between neighboring query terms and models this dependence by a Markov Random Field (MRF). The MRF in SDM defines a joint probability over a graph G
% , as shown in Figure \ref{fig:term_dependance},
whose nodes are document random variable ($\mathbf{D}$) and query terms $q_1, q_2, .., q_{|Q|}$. The edges between nodes represent the dependency between them, where a node is independent of all other nodes given its neighbors. 
% For example, $q_3$ is independent of $q_1$ given $\mathbf{D}$ and $q_2$.
The query document conditional relevance probability is defined via this joint probability factorized as follows:

\begin{align} 
        P_{\Lambda}(D|Q) &= \frac{P_{\Lambda}(Q,D)}{P_{\Lambda}(Q)} \stackrel{rank}{=} P_{\Lambda}(Q,D) \\ 
    & = \frac{1}{Z_\Lambda}\prod_{c\in C(G)}\psi(c,\Lambda) \stackrel{rank}{=} \sum_{c \in C(G)} \log \psi(c,\Lambda) \label{eq:sdm_mrf}
\end{align}
Here, $C(G)$ is the set of cliques in the graph, $\psi(c, \Lambda)$ is a potential function parameterized by $\Lambda$, and $Z_\Lambda$ is a normalization factor. SDM realizes the above formula by defining three %types of
potential functions:
\begin{itemize}[leftmargin=*]
    \item Exact individual term matching: 
    
    \resizebox{0.6\hsize}{!}{$\psi_{T}(q_i, D) = \lambda_T log\left[(1-\alpha_D)\frac{tf_{q_i, D}}{|D|} + \alpha_D \frac{cf_{q_i}}{|C|}\right]$}    
    \item  Exact n-gram/phrase matching:
    
    \resizebox{0.85\hsize}{!}{$\psi_{O}(q_i...q_{i+k}, D) = \lambda_O\log\left[(1-\alpha_D)\frac{tf_{\#1(q_i... q_{i+k}), D}}{|D|} + \alpha_D\frac{cf_{\#1(q_i...q_{i+k})}}{|C|}\right]$} 
    \item Exact multi-term proximity matching: \\
    \textit{(terms appear ordered/unordered within a window of size N)}
    
    \resizebox{0.9\hsize}{!}{$\psi_{U}(q_i...q_j, D) =  \lambda_U\log\left[(1-\alpha_D)\frac{tf_{ \#uwN(q_i...q_j), D}}{|D|} + \alpha_D\frac{cf_{ \#uwN(q_i...q_j)}}{|C|}\right]$}
    
    ($tf$ is term(s) frequency in $D$; $cf$ is term(s) frequency in the corpus $C$; $\lambda_{(s)}$ are learnable weights; $\alpha_D$ is a smoothing factor)
\end{itemize}
Plugging the above potential functions into Equation \ref{eq:sdm_mrf}, we get the following ranking function: 
\begin{multline}
P_{\Lambda}(D|Q) \stackrel{rank}{=} \sum_{\scriptscriptstyle q_i \in Q}\psi_{T}(q_i, D)  + \sum_{\scriptscriptstyle q_i...q_{i+k} \in Q}\psi_{O}(q_i...q_{i+k}, D) \\
  + \sum_{\scriptscriptstyle q_i...q_{j} \in Q}\psi_{U}(q_i...q_{j}, D)    \label{eq:ranking_function}      
\end{multline}
%\section{Methodology}
\section{Sequential Dependence for LSR} % Models for LSR}
% \begin{figure}
%     \centering
%     \includegraphics[trim=20 130 0 80,clip,width=1.1\linewidth]{images/term-dependence.pdf}
%     \caption{Sequential Dependence Assumption in MRF}
%     \label{fig:term_dependance}
%     \vspace{-0.5cm}
% \end{figure}
In this section, we will introduce SoftSDM and ExactSDM, which are adapted from the original SDM, for performing phrase and proximity matching with LSR. We formulate these two adaptations on top of the state-of-the-art Splade LSR encoder that utilizes BERT's masked language modeling head. While SoftSDM's formulation is tied to the MLM head, ExactSDM can be generalized to other LSR architectures that produce term weights differently. 

% In this work, we are interested in studying the potential of the SDM's proximity and phrase matching properties for adapting state-of-the-art learned sparse retrieval to long documents. To achieve this, we propose an adaptation of the SDM model to the MLM-heads used in state-of-the arts LSR methods. This adaptation naturally enables the SDM to exploit the soft-matching capability of the MLM head, we refer this adaptation as SoftSDM. 
\subsection{Query and document representations}
In order to measure the dependence between terms, it is necessary to track both term positions and weights. We utilize a modified version of Splade~\cite{nguyen2023unified} that enhances efficiency without compromising its effectiveness. It produces query term weights using an MLP layer (instead of MLM) on top of BERT's last hidden states, generating a sequence of term weights ($w_{q} = w_q^1, w_q^2 .. w_q^{|Q|}$) with $w_q^i$ representing the weight of the $i^{th}$ query token.

% $Q_{MLP}D_{MLM}$ uses the same MLM architecture as Splade on the document side.
%but we recommend storing the entire logit matrix ($W_{|D|x|V|}^D$) instead of just the max-pooling output, as Splade does.
% ^ do we need this detail for something later?
On the document side, the model's MLM document encoder produces for each document a logit matrix $\mathbf{W^D}$ where 
$W_{i,j}^D$ represents the translation score from the $i^{th}$ document term to the $j^{th}$ vocabulary item. This logit matrix provides access to term positions and preceding term-weight vectors. We sparsify $\mathbf{W^D}$ during training, enabling efficient storage via a sparse matrix format.

\subsection{Soft Sequential Dependence Model} \label{sec:potential_functions}
To adapt SoftSDM for use with the MLM head, 
%SoftSDM uses the same formulation as the original SDM, however
we introduce three new functions derived from the query and document representations described previously. These potential functions are defined as follows (\textit{with $v[q_i]$ used to denote the index of $q_i$ in the vocabulary}):
% \ay{Why do we need to say v[] instead of just qi?} 
% we want the index of q_i in the vocabulary ?? 
% but where is the index used? in the formulas 
% isn't it just a unique ID here? like qi 
% yes, it is unique 
% q_i is a token (text), we need the its index to get the weight of q_i in the logit matrix (W) 
% ah I get the notation now! Great 
% \ay{Can we add a sentence to each of these intuitively describing what is happening? I think the notation is hard to parse without this (especially for someone who doesn't already know what happens). Something like: for each query term, we find max matching expansion term within the ordered/unordered window.}
\begin{itemize}[leftmargin=*]
    \item \textbf{Soft individual term matching}: This potential function is equivalent to the max-aggregation followed by a dot product used in LSR methods (e.g., Splade).
    \begin{equation} \label{eq:st}
     \psi_{ST}(q_i, D) = \lambda_{ST}\max\limits_{1 \leq r \leq |D|}w_q^iW_{r,v[q_i]}^D
    \end{equation}
    \item \textbf{Soft n-gram/phrase matching}: Soft phrase similarity measures the likelihood of terms in one phrase translating to corresponding terms in another, considering the importance of each term. This function computes the maximum similarity between a query phrase and document phrases starting at every position $r$.
    % For each document phrase starting at $r$, it calculates the sum of translation score from all constituent term to the term at the same position in the query phrase.
    % Two phrases are similar if tokens in the same index are expansion terms of each other. The function compares  its similarity to every document phrases (starting at $r$) and return the max value.
    \begin{equation} \label{eq:so}
    \psi_{SO}(q_i...q_{i+k}, D) = \lambda_{SO}\max\limits_{{1 \leq r \leq |D|-k}}\sum_{l=0}^{k}w_q^{i+l}W_{r+l, v[q_{i+l}]}^D
    \end{equation}
    \item \textbf{Soft proximity matching}: This function approximates the maximum likelihood of translating terms within document windows of size p to a set of query terms regardless of the order, while also considering term importance.
    \begin{equation} \label{eq:su}
    \psi_{SU}(q_i...q_{i+k}, D) = \lambda_{SO}\max\limits_{{1 \leq r \leq |D|-p}} \sum_{h=i}^{i+k}w_q^h\left[\max\limits_{r \leq l < r+p} W_{l, v[q_h]}^D\right]  
    \end{equation}
\end{itemize}
% Given the definition of the above potential functions, the ranking formula in Equation \ref{eq:ranking_function} could be re-written as: 
% \begin{multline}
% P_{\Lambda}(D|Q) \stackrel{rank}{=} \sum_{\scriptscriptstyle q_i \in Q}\psi_{ST}(q_i, D)  + \sum_{\scriptscriptstyle q_i...q_{i+k} \in Q}\psi_{SO}(q_i...q_{i+k}, D) \\
%   + \sum_{\scriptscriptstyle q_i...q_{j} \in Q}\psi_{SU}(q_i...q_{i+k}, D)    \label{eq:ssdm_ranking_function}      
% \end{multline}
\subsubsection{Exact Sequential Dependence Model} ExactSDM uses the same potential function for individual term matching as SoftSDM (Equation \ref{eq:st}), which has been shown to benefit LSR~\cite{formal2021splade,formal2022distillation}. However, with ExactSDM, we are interested in evaluating the impact of soft phrase/proximity matching. It accomplishes this by allowing a document term to only translate to itself in the output logits and disabling document expansion. This change modifies the potential function formulas in Equation \ref{eq:so} and \ref{eq:su} by modifying the logit matrix to only contain self-translations.
% \ay{What do we mean by 'exchanging the new logit matrix'? I guess this is saying, setting all different term IDs to 0 translation score?}
ExactSDM is compatible with an MLM or MLP document encoder, making it applicable to additional LSR methods like uniCOIL~\cite{lin2021few}, DeepCT~\cite{dai2019context}, and DeepImpact~\cite{mallia2021learning}.
% , and DeepImpact~\cite{mallia2021learning}.
\vspace{-1.2em}
\section{Experiments}

\begin{table*}[ht]
        \caption{The results of baselines and SDM variants on MSMARCO Document and TREC Robust04.
        MRR and NDCG are @10. Recall is @1000. A $\dagger$ indicates $p < 0.05$ (paired t-test between a SDM method and Score-max with Bonferroni correction).}
        \label{tab:lsr_long_results}
        \vspace{-1em}
        \centering
        \captionsetup{justification=centering}
        \resizebox{\textwidth}{!}{
        \begin{tabular}{@{\extracolsep{1pt}}cccccccccccccccccc@{}}
        \toprule
        \toprule
        \multicolumn{2}{c}{\multirow{2}{*}{\textbf{\# segs}}} & \multicolumn{2}{c}{\textbf{ExactSDM}} & \multicolumn{2}{c}{\textbf{SoftSDM}} & \multicolumn{2}{c}{\textbf{Score-max}} & \multicolumn{2}{c}{\textbf{Rep-max}} & \multicolumn{2}{c}{\textbf{Rep(sc.)-sum}} & \multicolumn{2}{c}{\textbf{Rep(sc.)-mean}} & \multicolumn{2}{c}{\textbf{BM25-SDM}} &
        \multicolumn{2}{c}{\textbf{BM25-RM3}} \\ 
        \cline{3-4} \cline{5-6} \cline{7-8} \cline{9-10} \cline{11-12} \cline{13-14} \cline{15-16} \cline{17-18}
        \rule{0pt}{3ex}
        &  & MRR& R & MRR & R & MRR & R & MRR & R &MRR & R & MRR & R & MRR & R & MRR & R \\
        \midrule
        \multirow{5}{*}{\rotatebox[origin=c]{90}{\textbf{MSDoc Dev}}} & 1 & \textbf{37.08} & 95.49 & 36.98 & 95.49 & 36.63 & 95.49 & 36.63 & 95.49 & 36.63 & 95.49 & 36.63 & 95.49 & 26.09 & 89.91 & 21.13 & 91.39 \\ 
        & 2 & $^\dagger$37.45 & 96.51 & $^\dagger$\textbf{37.53} & 96.51 & 36.80 & 96.51 & 35.14 & 96.40 & 27.09 & 88.81 & 31.58 & 95.69 & 26.17 & 90.43 & 19.69 & 91.39 \\ 
        & 3 & $^\dagger$37.36 & 96.76 & $^\dagger$\textbf{37.41}& 96.76 & 36.72 & 96.76 & 33.91 & 96.34 &  19.71 & 75.24 & 29.02 & 95.13 & 25.91 & 90.45 & 18.83 &  90.78 \\ 
        & 4 &  $^\dagger$\textbf{37.03} & 96.71 & 36.80 & 96.71 & 32.72 & 96.71 & 14.52 & 96.30 & 36.36 & 65.51 & 27.94 & 94.43 & 25.58 & 90.33 & 17.94 & 90.20 \\ 
        & 5 & $^\dagger$\textbf{36.95} & 96.61 & $^\dagger$36.79 & 96.61 & 36.24 & 96.61 & 31.97 & 96.15 & 12.17 & 55.83 & 27.52 & 93.95 &  25.48 & 90.12 & 17.34 & 89.60 \\
        \midrule
        & & NDCG & R & NDCG & R & NDCG & R & NDCG & R & NDCG & R & NDCG & R & NDCG & R & NDCG & R \\
        \midrule
        \multirow{5}{*}{\rotatebox[origin=c]{90}{\textbf{Robust04}}} & 1 & $^\dagger$46.61 & 59.26 & $^\dagger$\textbf{46.65} & 59.26 & 44.88 & 59.26 & 44.88 & 59.26 & 44.88 & 59.26 & 44.88 & 59.26 & 39.63 & 54.84 & 42.05 & 61.76\\ 
        & 2 & $^\dagger$\textbf{47.98} & 64.28 & $^\dagger$47.94 & 64.28 & 45.71 & 64.28 &  43.88 & 63.37 & 42.23 & 51.90 & 44.32 & 59.88 & 42.04 & 60.79 & 43.04 &  68.21 \\ 
        & 3 & $^\dagger$\textbf{48.59} & 66.75 & $^\dagger$48.28 & 66.75 & 46.42 & 66.75 & 42.51 & 64.35 & 36.43 & 44.25 & 42.72 & 59.04 & 42.02 & 63.14 & 43.59 & 70.34\\ 
        & 4 & $^\dagger$\textbf{48.87} & 68.01 & $^\dagger$48.56 & 68.01 & 46.48 & 68.01 & 41.07 & 64.78 & 32.14 & 39.47 & 42.25 & 58.25 & 42.49 & 64.37 & 43.42 & 71.35\\ 
        & 5 & $^\dagger$\textbf{49.04} & 68.61 &  $^\dagger$48.65 & 68.61 & 46.37 & 68.61 & 40.50 & 64.86 & 28.41 & 36.02 & 41.77 & 58.10 & 42.46 & 64.63 & 44.02 & 70.97\\ 
        \bottomrule
        \end{tabular}}
    \end{table*}

\subsection{Datasets}
% \textbf{\underline{Datasets}}: 
Our experiments use two long-document benchmarks: MSMARCO Document and TREC Robust04. The MSMARCO Document dataset~\cite{msmarco} includes 3.2 million documents, 5.2K dev queries, and 367K queries in the train split. TREC Robust04~\cite{robust04} consists of approximately 0.5 million news articles and 250 queries, with each query containing three fields: title, description, and narrative. In this work, we use only the description field, which contains a natural language version of the query. We access these datasets using  ir\_datasets~\cite{macavaney:sigir2021-irds}.
\vspace{-2.5em}
\subsection{Experimental settings}    
% \noindent\textbf{\underline{}}:  
We utilize the Splade-based LSR architecture using DistilBERT~\cite{sanh2019distilbert} with a maximum input length of 512 tokens for all experiments. This architecture was trained on the MS-MARCO passage dataset using hard negatives and distillation from a Cross-Encoder~\cite{reimers2019sentence}. To handle long documents, we divided them into sentences and grouped sentences into segments of up to 400 tokens, with longer sentences considered as a single segment. Document representations or scores were derived from the segments' representations/scores generated by the Splade model, which achieved a Passage MRR@10 of 38.51 when previously trained on the MS-MARCO passage dataset~\cite{msmarco} (with no fine-tuning on the document dataset).
    
    We explore various aggregation approaches, including Rep-max, Score-max, Rep(score)-sum, and Rep(score)-mean, that describe how segment-level term scores are treated to arrive at a document-level score.
    %Each aggregation approach is responsible for assigning a weight to a vocabulary term.
    % \ay{We don't explain these in detail earlier, so we should make sure they're clear here.}
    The representation aggregation methods (e.g., Rep-max) operate on the sparse vector corresponding to each segment, whereas the score aggregation methods (e.g., Score-max) operate on relevance scores assigned to each segment.
    With Rep-max, a document-level sparse vector is computed by taking the maximum weight for each vocabulary term across all document segments. This document-level vector is then used to compute the relevance score by taking the dot product with the query vector.
    Score-max calculates a relevance score between each segment and the query, and then returns the maximum segment relevance score as the document-level relevance score.
    %Rep-max assigns the maximum weight to each vocabulary item across document segments, while Score-max calculates the relevance (dot product) between each segment and the query and returned the maximum score.
    Rep-sum and Rep-mean replace Rep-max's max pooling with sum pooling or mean pooling across segments.
    Due to the distributive property of dot products, sum and mean pooling return identical results regardless of whether they operate on sparse vectors (representations) or scores. 
    Thus, we refer to these as Rep(score)-sum and Rep(score)-mean.
    %Rep(score)-sum performed per-term or score pooling across different segments, and Rep(score)-mean took the result of sum pooling and divided it by the number of segments.

    SoftSDM and ExactSDM were evaluated using the same query and document segment representations as the above approaches, but with additional positional information. Score-max can be viewed as a special case of SoftSDM with soft proximity matching within the segment windows.
    % \ay{I think people will get confused about whether the above aggregation choices still affect Soft/ExactSDM. Is there an easy way to explain that *SDM is a complete replacement for the above choices? (Rep-max, Rep-sum etc)} 
    % \tn{Can we say like Score-max is a special case of proximity's potential function when the window_size = 400?}
    When using these approaches, we fine-tune the three weighting factors of the potential functions using the MSMARCO Document training triplets obtained from~\citet{boytsov2022understanding}. We also assess the generalizability of these weighting factors on TREC Robust04 without any further fine-tuning.
    \vspace{-1em}
\subsection{Results and Discussion}
In our experiments, we compare existing aggregation methods with our proposed SDM for LSR variants as the number of segments in documents increases.
%We first look at the performance of existing long-document aggregation methods evaluated with LSR.

We first look at the performance of existing long-document aggregation methods evaluated with LSR. 
Figure \ref{fig:lsr_long_figure} illustrates the results of segment aggregation techniques with respect to the number of segments, with a detailed breakdown in Table \ref{tab:lsr_long_results}. We observe a strong downward trend with existing approaches as more document segments are considered, with the exception of Score-max on both the MSMARCO Document and Robust04 datasets.
        
The approach of summing segment scores or representations, Rep(Score)-sum, suffers from the most severe decrease in performance on the MRR@10 and NDCG@10 measures. The MRR@10/ NDCG@10 on MSMARCO Document/Robust04 drops significantly from 36.63/44.88 when using the first segment to worse than BM25 SDM/RM3 when 10 segments are used. The gap becomes even larger (not shown in the Table) when all segments are used, while BM25's scores only slightly decrease. This issue can be attributed to the longer and noisier texts resulting from using more segments, which leads to the addition of more non-relevant terms to the document's representation. Moreover, sum aggregation is inherently biased towards longer documents; less relevant documents may receive a higher accumulated score than shorter but more relevant ones.
        
The Rep(score)-mean aggregation method corrects for this bias by dividing the sum by the number of segments, thus exhibiting a clear recovery on both datasets. We also note the competitiveness between Rep(score)-mean and Rep-max. On MSMARCO Document, Rep-max consistently outperforms the mean aggregation, but the opposite is true for Robust04. 
In contrast with prior results on Cross-Encoders~\cite{li2020parade}, where representation aggregation methods are preferable, the Score-max approach outperforms all other aggregation methods and is more robust than the max representation pooling. The MRR@10 of Score-max slightly drops on MSMARCO Document, while its NDCG@10 on Robust04 goes up initially, up to the third segment, and becomes stable after that. The local scoring capability of Score-max likely enables it to avoid noise accumulation.
This discrepancy between performance with LSR and with Cross-Encoders may be due to the fact that Cross-Encoders can effectively ignore non-relevant segments, because they know both the query and the document at the time of encoding. LSR methods must encode documents without any knowledge of the query.

The performance of Score-max in the face of a downtrend suggests that proximity is crucial for effectively scoring long documents using LSR. This observation leads to the ExactSDM and SoftSDM models, both of which are adapted from the well-known SDM model from the pre-neural era. As shown in the two leftmost blocks of Table \ref{tab:lsr_long_results}, both ExactSDM and SoftSDM consistently outperform previous aggregation approaches and BM25 RM3/SDM on both datasets, regardless of varying document lengths. Overall, ExactSDM and SoftSDM perform competitively on the two datasets.

On MSMARCO Document, the MRR@10 of both SoftSDM and ExactSDM slightly increases on the first two or three segments, but starts to fluctuate after that. In comparison with the previous best pooling method, Score-max, both SDM variants achieve better MRR@10, with the improvement ranging from 1.0\% to 2.0\%. On Robust04 (zero-shot), the improvement over Score-max is even greater, ranging from 3.9\% to 5.8\%. Additionally, the improvement slightly increases when more segments are used, demonstrating the ability of SDM variants to exploit long context.

% We observe a positive, weak correlation between the gain over Score-max of SoftSDM and ExactSDM and the number of segments, which is partly due to the decline of Score-max on MSMARCO doc. However, it also demonstrates the robustness of SoftSDM and ExactSDM in exploiting long documents' context.
Comparing SoftSDM and ExactSDM, we see that ExactSDM slightly outperforms SoftSDM, making it a great choice for modeling term dependence with LSR. ExactSDM does not necessarily rely on the MLM's logits, which means it is compatible with other LSR methods not using the MLM encoders. 
% In a pilot study, we found that combining the potential functions of both SoftSDM and ExactSDM does not improve performance. 

Regarding the optimal phrase and proximity window size, we found that using two-term phrases (bi-grams) and a proximity window of size 8 often returns higher results on MSMARCO Document, which is consistent with the recommendations in~\cite{metzler2005modeling,bendersky2010learning}. While this setting generalizes well to TREC Robust04 with ExactSDM, SoftSDM needs longer phrases (5 grams) and a longer proximity window (size=10) for better zero-shot performance.  
\vspace{-0.5em}
\section{Conclusion}
Our work explores techniques for adapting learned sparse retrieval to long documents. We find that the max-score aggregation approach is robust to varying document lengths, leading to our SoftSDM and Exact SDM adaptations of the SDM model that outperform existing approaches. Surprisingly, SoftSDM's soft-matching ability does not outperform ExactSDM, indicating that it may not be necessary for modeling term dependence. 

% Additionally, we plan to investigate if the MLM-based LSR encoder retains positional information of expanded terms in future work.

%TODO HI ack for camera ready

\bibliographystyle{ACM-Reference-Format}
\balance
\bibliography{sigir23-main}


\begin{thebibliography}{21}

%%% the Web address.

\ifx \showCODEN    \undefined \def \showCODEN     #1{\unskip}     \fi
\ifx \showDOI      \undefined \def \showDOI       #1{#1}\fi
\ifx \showISBNx    \undefined \def \showISBNx     #1{\unskip}     \fi
\ifx \showISBNxiii \undefined \def \showISBNxiii  #1{\unskip}     \fi
\ifx \showISSN     \undefined \def \showISSN      #1{\unskip}     \fi
\ifx \showLCCN     \undefined \def \showLCCN      #1{\unskip}     \fi
\ifx \shownote     \undefined \def \shownote      #1{#1}          \fi
\ifx \showarticletitle \undefined \def \showarticletitle #1{#1}   \fi
\ifx \showURL      \undefined \def \showURL       {\relax}        \fi
% invisible to TeX
\providecommand\bibfield[2]{#2}
\providecommand\bibinfo[2]{#2}
\providecommand\natexlab[1]{#1}
\providecommand\showeprint[2][]{arXiv:#2}

\bibitem[Bendersky et~al\mbox{.}(2010)]%
        {bendersky2010learning}
\bibfield{author}{\bibinfo{person}{Michael Bendersky}, \bibinfo{person}{Donald
  Metzler}, {and} \bibinfo{person}{W.~Bruce Croft}.}
  \bibinfo{year}{2010}\natexlab{}.
\newblock \showarticletitle{Learning concept importance using a weighted
  dependence model}. In \bibinfo{booktitle}{\emph{Proceedings of the Third
  International Conference on Web Search and Web Data Mining, {WSDM} 2010, New
  York, NY, USA, February 4-6, 2010}},
  \bibfield{editor}{\bibinfo{person}{Brian~D. Davison},
  \bibinfo{person}{Torsten Suel}, \bibinfo{person}{Nick Craswell}, {and}
  \bibinfo{person}{Bing Liu}} (Eds.). \bibinfo{publisher}{{ACM}},
  \bibinfo{pages}{31--40}.
\newblock
\urldef\tempurl%
\url{https://doi.org/10.1145/1718487.1718492}
\showDOI{\tempurl}


\bibitem[Boytsov et~al\mbox{.}(2022)]%
        {boytsov2022understanding}
\bibfield{author}{\bibinfo{person}{Leonid Boytsov}, \bibinfo{person}{Tianyi
  Lin}, \bibinfo{person}{Fangwei Gao}, \bibinfo{person}{Yutian Zhao},
  \bibinfo{person}{Jeffrey Huang}, {and} \bibinfo{person}{Eric Nyberg}.}
  \bibinfo{year}{2022}\natexlab{}.
\newblock \showarticletitle{Understanding Performance of Long-Document Ranking
  Models through Comprehensive Evaluation and Leaderboarding}.
\newblock \bibinfo{journal}{\emph{CoRR}}  \bibinfo{volume}{abs/2207.01262}
  (\bibinfo{year}{2022}).
\newblock
\urldef\tempurl%
\url{https://doi.org/10.48550/arXiv.2207.01262}
\showDOI{\tempurl}
\showeprint[arXiv]{2207.01262}


\bibitem[Dai and Callan(2019)]%
        {dai2019context}
\bibfield{author}{\bibinfo{person}{Zhuyun Dai} {and} \bibinfo{person}{Jamie
  Callan}.} \bibinfo{year}{2019}\natexlab{}.
\newblock \showarticletitle{Context-Aware Sentence/Passage Term Importance
  Estimation For First Stage Retrieval}.
\newblock \bibinfo{journal}{\emph{CoRR}}  \bibinfo{volume}{abs/1910.10687}
  (\bibinfo{year}{2019}).
\newblock
\showeprint[arXiv]{1910.10687}
\urldef\tempurl%
\url{http://arxiv.org/abs/1910.10687}
\showURL{%
\tempurl}


\bibitem[Formal et~al\mbox{.}(2022)]%
        {formal2022distillation}
\bibfield{author}{\bibinfo{person}{Thibault Formal}, \bibinfo{person}{Carlos
  Lassance}, \bibinfo{person}{Benjamin Piwowarski}, {and}
  \bibinfo{person}{St{\'{e}}phane Clinchant}.} \bibinfo{year}{2022}\natexlab{}.
\newblock \showarticletitle{From Distillation to Hard Negative Sampling: Making
  Sparse Neural {IR} Models More Effective}. In
  \bibinfo{booktitle}{\emph{{SIGIR} '22: The 45th International {ACM} {SIGIR}
  Conference on Research and Development in Information Retrieval, Madrid,
  Spain, July 11 - 15, 2022}}, \bibfield{editor}{\bibinfo{person}{Enrique
  Amig{\'{o}}}, \bibinfo{person}{Pablo Castells}, \bibinfo{person}{Julio
  Gonzalo}, \bibinfo{person}{Ben Carterette}, \bibinfo{person}{J.~Shane
  Culpepper}, {and} \bibinfo{person}{Gabriella Kazai}} (Eds.).
  \bibinfo{publisher}{{ACM}}, \bibinfo{pages}{2353--2359}.
\newblock
\urldef\tempurl%
\url{https://doi.org/10.1145/3477495.3531857}
\showDOI{\tempurl}


\bibitem[Formal et~al\mbox{.}(2021)]%
        {formal2021splade}
\bibfield{author}{\bibinfo{person}{Thibault Formal}, \bibinfo{person}{Benjamin
  Piwowarski}, {and} \bibinfo{person}{St{\'{e}}phane Clinchant}.}
  \bibinfo{year}{2021}\natexlab{}.
\newblock \showarticletitle{{SPLADE:} Sparse Lexical and Expansion Model for
  First Stage Ranking}. In \bibinfo{booktitle}{\emph{{SIGIR} '21: The 44th
  International {ACM} {SIGIR} Conference on Research and Development in
  Information Retrieval, Virtual Event, Canada, July 11-15, 2021}},
  \bibfield{editor}{\bibinfo{person}{Fernando Diaz}, \bibinfo{person}{Chirag
  Shah}, \bibinfo{person}{Torsten Suel}, \bibinfo{person}{Pablo Castells},
  \bibinfo{person}{Rosie Jones}, {and} \bibinfo{person}{Tetsuya Sakai}} (Eds.).
  \bibinfo{publisher}{{ACM}}, \bibinfo{pages}{2288--2292}.
\newblock
\urldef\tempurl%
\url{https://doi.org/10.1145/3404835.3463098}
\showDOI{\tempurl}


\bibitem[Huston and Croft(2014)]%
        {huston2014comparison}
\bibfield{author}{\bibinfo{person}{Samuel~J. Huston} {and}
  \bibinfo{person}{W.~Bruce Croft}.} \bibinfo{year}{2014}\natexlab{}.
\newblock \showarticletitle{A Comparison of Retrieval Models using Term
  Dependencies}. In \bibinfo{booktitle}{\emph{Proceedings of the 23rd {ACM}
  International Conference on Conference on Information and Knowledge
  Management, {CIKM} 2014, Shanghai, China, November 3-7, 2014}},
  \bibfield{editor}{\bibinfo{person}{Jianzhong Li},
  \bibinfo{person}{Xiaoyang~Sean Wang}, \bibinfo{person}{Minos~N. Garofalakis},
  \bibinfo{person}{Ian Soboroff}, \bibinfo{person}{Torsten Suel}, {and}
  \bibinfo{person}{Min Wang}} (Eds.). \bibinfo{publisher}{{ACM}},
  \bibinfo{pages}{111--120}.
\newblock
\urldef\tempurl%
\url{https://doi.org/10.1145/2661829.2661894}
\showDOI{\tempurl}


\bibitem[Li et~al\mbox{.}(2020)]%
        {li2020parade}
\bibfield{author}{\bibinfo{person}{Canjia Li}, \bibinfo{person}{Andrew Yates},
  \bibinfo{person}{Sean MacAvaney}, \bibinfo{person}{Ben He}, {and}
  \bibinfo{person}{Yingfei Sun}.} \bibinfo{year}{2020}\natexlab{}.
\newblock \showarticletitle{{PARADE:} Passage Representation Aggregation for
  Document Reranking}.
\newblock \bibinfo{journal}{\emph{CoRR}}  \bibinfo{volume}{abs/2008.09093}
  (\bibinfo{year}{2020}).
\newblock
\showeprint[arXiv]{2008.09093}
\urldef\tempurl%
\url{https://arxiv.org/abs/2008.09093}
\showURL{%
\tempurl}


\bibitem[Lin(2021)]%
        {lin2022proposed}
\bibfield{author}{\bibinfo{person}{Jimmy Lin}.}
  \bibinfo{year}{2021}\natexlab{}.
\newblock \showarticletitle{A proposed conceptual framework for a
  representational approach to information retrieval}.
\newblock \bibinfo{journal}{\emph{{SIGIR} Forum}} \bibinfo{volume}{55},
  \bibinfo{number}{2}, \bibinfo{pages}{4:1--4:29}.
\newblock
\urldef\tempurl%
\url{https://doi.org/10.1145/3527546.3527552}
\showDOI{\tempurl}


\bibitem[Lin and Ma(2021)]%
        {lin2021few}
\bibfield{author}{\bibinfo{person}{Jimmy Lin} {and} \bibinfo{person}{Xueguang
  Ma}.} \bibinfo{year}{2021}\natexlab{}.
\newblock \showarticletitle{A Few Brief Notes on DeepImpact, COIL, and a
  Conceptual Framework for Information Retrieval Techniques}.
\newblock \bibinfo{journal}{\emph{CoRR}}  \bibinfo{volume}{abs/2106.14807}
  (\bibinfo{year}{2021}).
\newblock
\showeprint[arXiv]{2106.14807}
\urldef\tempurl%
\url{https://arxiv.org/abs/2106.14807}
\showURL{%
\tempurl}


\bibitem[MacAvaney et~al\mbox{.}(2020)]%
        {macavaney2020expansion}
\bibfield{author}{\bibinfo{person}{Sean MacAvaney},
  \bibinfo{person}{Franco~Maria Nardini}, \bibinfo{person}{Raffaele Perego},
  \bibinfo{person}{Nicola Tonellotto}, \bibinfo{person}{Nazli Goharian}, {and}
  \bibinfo{person}{Ophir Frieder}.} \bibinfo{year}{2020}\natexlab{}.
\newblock \showarticletitle{Expansion via Prediction of Importance with
  Contextualization}. In \bibinfo{booktitle}{\emph{Proceedings of the 43rd
  International {ACM} {SIGIR} conference on research and development in
  Information Retrieval, {SIGIR} 2020, Virtual Event, China, July 25-30,
  2020}}, \bibfield{editor}{\bibinfo{person}{Jimmy~X. Huang},
  \bibinfo{person}{Yi~Chang}, \bibinfo{person}{Xueqi Cheng},
  \bibinfo{person}{Jaap Kamps}, \bibinfo{person}{Vanessa Murdock},
  \bibinfo{person}{Ji{-}Rong Wen}, {and} \bibinfo{person}{Yiqun Liu}} (Eds.).
  \bibinfo{publisher}{{ACM}}, \bibinfo{pages}{1573--1576}.
\newblock
\urldef\tempurl%
\url{https://doi.org/10.1145/3397271.3401262}
\showDOI{\tempurl}


\bibitem[MacAvaney et~al\mbox{.}(2019)]%
        {macavaney:sigir2019-cedr}
\bibfield{author}{\bibinfo{person}{Sean MacAvaney}, \bibinfo{person}{Andrew
  Yates}, \bibinfo{person}{Arman Cohan}, {and} \bibinfo{person}{Nazli
  Goharian}.} \bibinfo{year}{2019}\natexlab{}.
\newblock \showarticletitle{{CEDR:} Contextualized Embeddings for Document
  Ranking}. In \bibinfo{booktitle}{\emph{Proceedings of the 42nd International
  {ACM} {SIGIR} Conference on Research and Development in Information
  Retrieval, {SIGIR} 2019, Paris, France, July 21-25, 2019}},
  \bibfield{editor}{\bibinfo{person}{Benjamin Piwowarski}, \bibinfo{person}{Max
  Chevalier}, \bibinfo{person}{{\'{E}}ric Gaussier}, \bibinfo{person}{Yoelle
  Maarek}, \bibinfo{person}{Jian{-}Yun Nie}, {and} \bibinfo{person}{Falk
  Scholer}} (Eds.). \bibinfo{publisher}{{ACM}}, \bibinfo{pages}{1101--1104}.
\newblock
\urldef\tempurl%
\url{https://doi.org/10.1145/3331184.3331317}
\showDOI{\tempurl}


\bibitem[MacAvaney et~al\mbox{.}(2021)]%
        {macavaney:sigir2021-irds}
\bibfield{author}{\bibinfo{person}{Sean MacAvaney}, \bibinfo{person}{Andrew
  Yates}, \bibinfo{person}{Sergey Feldman}, \bibinfo{person}{Doug Downey},
  \bibinfo{person}{Arman Cohan}, {and} \bibinfo{person}{Nazli Goharian}.}
  \bibinfo{year}{2021}\natexlab{}.
\newblock \showarticletitle{Simplified Data Wrangling with ir{\_}datasets}. In
  \bibinfo{booktitle}{\emph{{SIGIR} '21: The 44th International {ACM} {SIGIR}
  Conference on Research and Development in Information Retrieval, Virtual
  Event, Canada, July 11-15, 2021}},
  \bibfield{editor}{\bibinfo{person}{Fernando Diaz}, \bibinfo{person}{Chirag
  Shah}, \bibinfo{person}{Torsten Suel}, \bibinfo{person}{Pablo Castells},
  \bibinfo{person}{Rosie Jones}, {and} \bibinfo{person}{Tetsuya Sakai}} (Eds.).
  \bibinfo{publisher}{{ACM}}, \bibinfo{pages}{2429--2436}.
\newblock
\urldef\tempurl%
\url{https://doi.org/10.1145/3404835.3463254}
\showDOI{\tempurl}


\bibitem[Mallia et~al\mbox{.}(2021)]%
        {mallia2021learning}
\bibfield{author}{\bibinfo{person}{Antonio Mallia}, \bibinfo{person}{Omar
  Khattab}, \bibinfo{person}{Torsten Suel}, {and} \bibinfo{person}{Nicola
  Tonellotto}.} \bibinfo{year}{2021}\natexlab{}.
\newblock \showarticletitle{Learning Passage Impacts for Inverted Indexes}. In
  \bibinfo{booktitle}{\emph{{SIGIR} '21: The 44th International {ACM} {SIGIR}
  Conference on Research and Development in Information Retrieval, Virtual
  Event, Canada, July 11-15, 2021}},
  \bibfield{editor}{\bibinfo{person}{Fernando Diaz}, \bibinfo{person}{Chirag
  Shah}, \bibinfo{person}{Torsten Suel}, \bibinfo{person}{Pablo Castells},
  \bibinfo{person}{Rosie Jones}, {and} \bibinfo{person}{Tetsuya Sakai}} (Eds.).
  \bibinfo{publisher}{{ACM}}, \bibinfo{pages}{1723--1727}.
\newblock
\urldef\tempurl%
\url{https://doi.org/10.1145/3404835.3463030}
\showDOI{\tempurl}


\bibitem[Metzler and Croft(2005)]%
        {metzler2005modeling}
\bibfield{author}{\bibinfo{person}{Donald Metzler} {and}
  \bibinfo{person}{W.~Bruce Croft}.} \bibinfo{year}{2005}\natexlab{}.
\newblock \showarticletitle{A Markov random field model for term dependencies}.
  In \bibinfo{booktitle}{\emph{{SIGIR} 2005: Proceedings of the 28th Annual
  International {ACM} {SIGIR} Conference on Research and Development in
  Information Retrieval, Salvador, Brazil, August 15-19, 2005}},
  \bibfield{editor}{\bibinfo{person}{Ricardo~A. Baeza{-}Yates},
  \bibinfo{person}{Nivio Ziviani}, \bibinfo{person}{Gary Marchionini},
  \bibinfo{person}{Alistair Moffat}, {and} \bibinfo{person}{John Tait}} (Eds.).
  \bibinfo{publisher}{{ACM}}, \bibinfo{pages}{472--479}.
\newblock
\urldef\tempurl%
\url{https://doi.org/10.1145/1076034.1076115}
\showDOI{\tempurl}


\bibitem[Nguyen et~al\mbox{.}(2023)]%
        {nguyen2023unified}
\bibfield{author}{\bibinfo{person}{Thong Nguyen}, \bibinfo{person}{Sean
  MacAvaney}, {and} \bibinfo{person}{Andrew Yates}.}
  \bibinfo{year}{2023}\natexlab{}.
\newblock \showarticletitle{A Unified Framework for Learned Sparse Retrieval}.
  In \bibinfo{booktitle}{\emph{Advances in Information Retrieval - 45th
  European Conference on Information Retrieval, {ECIR} 2023, Dublin, Ireland,
  April 2-6, 2023, Proceedings, Part {III}}} \emph{(\bibinfo{series}{Lecture
  Notes in Computer Science}, Vol.~\bibinfo{volume}{13982})},
  \bibfield{editor}{\bibinfo{person}{Jaap Kamps}, \bibinfo{person}{Lorraine
  Goeuriot}, \bibinfo{person}{Fabio Crestani}, \bibinfo{person}{Maria Maistro},
  \bibinfo{person}{Hideo Joho}, \bibinfo{person}{Brian Davis},
  \bibinfo{person}{Cathal Gurrin}, \bibinfo{person}{Udo Kruschwitz}, {and}
  \bibinfo{person}{Annalina Caputo}} (Eds.). \bibinfo{publisher}{Springer},
  \bibinfo{pages}{101--116}.
\newblock
\urldef\tempurl%
\url{https://doi.org/10.1007/978-3-031-28241-6\_7}
\showDOI{\tempurl}


\bibitem[Nguyen et~al\mbox{.}(2016)]%
        {msmarco}
\bibfield{author}{\bibinfo{person}{Tri Nguyen}, \bibinfo{person}{Mir
  Rosenberg}, \bibinfo{person}{Xia Song}, \bibinfo{person}{Jianfeng Gao},
  \bibinfo{person}{Saurabh Tiwary}, \bibinfo{person}{Rangan Majumder}, {and}
  \bibinfo{person}{Li Deng}.} \bibinfo{year}{2016}\natexlab{}.
\newblock \showarticletitle{{MS} {MARCO:} {A} Human Generated MAchine Reading
  COmprehension Dataset}. In \bibinfo{booktitle}{\emph{Proceedings of the
  Workshop on Cognitive Computation: Integrating neural and symbolic approaches
  2016 co-located with the 30th Annual Conference on Neural Information
  Processing Systems {(NIPS} 2016), Barcelona, Spain, December 9, 2016}}
  \emph{(\bibinfo{series}{{CEUR} Workshop Proceedings},
  Vol.~\bibinfo{volume}{1773})},
  \bibfield{editor}{\bibinfo{person}{Tarek~Richard Besold},
  \bibinfo{person}{Antoine Bordes}, \bibinfo{person}{Artur~S. d'Avila Garcez},
  {and} \bibinfo{person}{Greg Wayne}} (Eds.). \bibinfo{publisher}{CEUR-WS.org}.
\newblock
\urldef\tempurl%
\url{https://ceur-ws.org/Vol-1773/CoCoNIPS\_2016\_paper9.pdf}
\showURL{%
\tempurl}


\bibitem[Reimers and Gurevych(2019)]%
        {reimers2019sentence}
\bibfield{author}{\bibinfo{person}{Nils Reimers} {and} \bibinfo{person}{Iryna
  Gurevych}.} \bibinfo{year}{2019}\natexlab{}.
\newblock \showarticletitle{Sentence-BERT: Sentence Embeddings using Siamese
  BERT-Networks}. In \bibinfo{booktitle}{\emph{Proceedings of the 2019
  Conference on Empirical Methods in Natural Language Processing and the 9th
  International Joint Conference on Natural Language Processing, {EMNLP-IJCNLP}
  2019, Hong Kong, China, November 3-7, 2019}},
  \bibfield{editor}{\bibinfo{person}{Kentaro Inui}, \bibinfo{person}{Jing
  Jiang}, \bibinfo{person}{Vincent Ng}, {and} \bibinfo{person}{Xiaojun Wan}}
  (Eds.). \bibinfo{publisher}{Association for Computational Linguistics},
  \bibinfo{pages}{3980--3990}.
\newblock
\urldef\tempurl%
\url{https://doi.org/10.18653/v1/D19-1410}
\showDOI{\tempurl}


\bibitem[Sanh et~al\mbox{.}(2019)]%
        {sanh2019distilbert}
\bibfield{author}{\bibinfo{person}{Victor Sanh}, \bibinfo{person}{Lysandre
  Debut}, \bibinfo{person}{Julien Chaumond}, {and} \bibinfo{person}{Thomas
  Wolf}.} \bibinfo{year}{2019}\natexlab{}.
\newblock \showarticletitle{DistilBERT, a distilled version of {BERT:} smaller,
  faster, cheaper and lighter}.
\newblock \bibinfo{journal}{\emph{CoRR}}  \bibinfo{volume}{abs/1910.01108}
  (\bibinfo{year}{2019}).
\newblock
\showeprint[arXiv]{1910.01108}
\urldef\tempurl%
\url{http://arxiv.org/abs/1910.01108}
\showURL{%
\tempurl}


\bibitem[Tao and Zhai(2007)]%
        {Tao2007AnEO}
\bibfield{author}{\bibinfo{person}{Tao Tao} {and} \bibinfo{person}{ChengXiang
  Zhai}.} \bibinfo{year}{2007}\natexlab{}.
\newblock \showarticletitle{An exploration of proximity measures in information
  retrieval}. In \bibinfo{booktitle}{\emph{{SIGIR} 2007: Proceedings of the
  30th Annual International {ACM} {SIGIR} Conference on Research and
  Development in Information Retrieval, Amsterdam, The Netherlands, July 23-27,
  2007}}, \bibfield{editor}{\bibinfo{person}{Wessel Kraaij},
  \bibinfo{person}{Arjen~P. de~Vries}, \bibinfo{person}{Charles L.~A. Clarke},
  \bibinfo{person}{Norbert Fuhr}, {and} \bibinfo{person}{Noriko Kando}} (Eds.).
  \bibinfo{publisher}{{ACM}}, \bibinfo{pages}{295--302}.
\newblock
\urldef\tempurl%
\url{https://doi.org/10.1145/1277741.1277794}
\showDOI{\tempurl}


\bibitem[Voorhees(2004)]%
        {robust04}
\bibfield{author}{\bibinfo{person}{Ellen~M. Voorhees}.}
  \bibinfo{year}{2004}\natexlab{}.
\newblock \showarticletitle{Overview of the {TREC} 2004 Robust Track}. In
  \bibinfo{booktitle}{\emph{Proceedings of The Thirteenth Text REtrieval
  Conference, {TREC} 2004, Gaithersburg, Maryland, USA}}
  \emph{(\bibinfo{series}{{NIST} Special Publication})},
  \bibfield{editor}{\bibinfo{person}{Ellen~M. Voorhees} {and}
  \bibinfo{person}{Lori~P. Buckland}} (Eds.). \bibinfo{publisher}{National
  Institute of Standards and Technology {(NIST)}}.
\newblock
\urldef\tempurl%
\url{https://trec.nist.gov/pubs/trec13/papers/ROBUST.OVERVIEW.pdf}
\showURL{%
\tempurl}


\bibitem[Zhang et~al\mbox{.}(2021)]%
        {zhang2021comparing}
\bibfield{author}{\bibinfo{person}{Xinyu Zhang}, \bibinfo{person}{Andrew
  Yates}, {and} \bibinfo{person}{Jimmy Lin}.} \bibinfo{year}{2021}\natexlab{}.
\newblock \showarticletitle{Comparing Score Aggregation Approaches for Document
  Retrieval with Pretrained Transformers}. In
  \bibinfo{booktitle}{\emph{Advances in Information Retrieval - 43rd European
  Conference on {IR} Research, {ECIR} 2021, Virtual Event, March 28 - April 1,
  2021, Proceedings, Part {II}}} \emph{(\bibinfo{series}{Lecture Notes in
  Computer Science}, Vol.~\bibinfo{volume}{12657})},
  \bibfield{editor}{\bibinfo{person}{Djoerd Hiemstra},
  \bibinfo{person}{Marie{-}Francine Moens}, \bibinfo{person}{Josiane Mothe},
  \bibinfo{person}{Raffaele Perego}, \bibinfo{person}{Martin Potthast}, {and}
  \bibinfo{person}{Fabrizio Sebastiani}} (Eds.). \bibinfo{publisher}{Springer},
  \bibinfo{pages}{150--163}.
\newblock
\urldef\tempurl%
\url{https://doi.org/10.1007/978-3-030-72240-1\_11}
\showDOI{\tempurl}


\end{thebibliography}

\end{document}